\documentclass[sigconf]{acmart}

\AtBeginDocument{%
  \providecommand\BibTeX{{%
    Bib\TeX}}}
\copyrightyear{2025}
\acmYear{2025}
\setcopyright{rightsretained}
\acmConference[FPGA '25]{Proceedings of the 2025 ACM/SIGDA International Symposium on Field Programmable Gate Arrays}{February 27--March 1, 2025}{Monterey, CA, USA}
\acmBooktitle{Proceedings of the 2025 ACM/SIGDA International Symposium on Field Programmable Gate Arrays (FPGA '25), February 27--March 1, 2025, Monterey, CA, USA}
\acmDOI{10.1145/3706628.3708823}
\acmISBN{979-8-4007-1396-5/25/02}

\makeatletter
\gdef\@copyrightpermission{
   \begin{minipage}{0.2\columnwidth}
     \href{https://creativecommons.org/licenses/by-nc-sa/4.0/}{\includegraphics[width=0.90\textwidth]{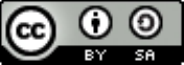}}
   \end{minipage}\hfill
   \begin{minipage}{0.8\columnwidth}
     \href{https://creativecommons.org/licenses/by-nc-sa/4.0/}{This work is licensed under a Creative Commons Attribution-ShareAlike International 4.0 License.}
   \end{minipage}
   \vspace{5pt}
}
\makeatother

\usepackage{algorithmic}
\usepackage{graphicx}
\usepackage{textcomp}
\usepackage{multirow}
\usepackage{xcolor}
\usepackage{pgfplots}
\usepackage{booktabs}
\usepackage{subcaption}
\pgfplotsset{compat=1.18}
\usepackage{soul}
\usepackage{array}
\usepackage{setspace}

\usepackage[ruled, linesnumbered, vlined, longend]{algorithm2e}
\usepackage{setspace}
\SetAlFnt{\footnotesize}
\SetAlCapNameFnt{\small}
\SetAlCapFnt{\small}
\IncMargin{3pt}
\SetAlgoNlRelativeSize{-0.5}
\SetKwProg{Def}{def}{$\,$:}{}
\SetKwProg{Func}{Func}{$\,$:}{}
\SetKwInOut{Input}{Input}
\SetKwInOut{Output}{Output}

\newcommand \jsctwolmaxred{36\%}
\newcommand \jscfivelmaxred{39\%}
\newcommand \mnistmaxred{11\%}

\def\BibTeX{{\rm B\kern-.05em{\sc i\kern-.025em b}\kern-.08em
    T\kern-.1667em\lower.7ex\hbox{E}\kern-.125emX}}

\begin{document}

\title{ReducedLUT: Table Decomposition with ``Don't Care'' Conditions}

\author{Oliver Cassidy}
\email{olly.cassidy23@imperial.ac.uk}
\affiliation{%
  \institution{Imperial College London}
  \city{London}
  \country{UK}
} 

\author{Marta Andronic}
\email{marta.andronic18@imperial.ac.uk}
\affiliation{%
  \institution{Imperial College London}
  \city{London}
  \country{UK}
}

\author{Samuel Coward}
\email{s.coward21@imperial.ac.uk}
\affiliation{%
  \institution{Intel and Imperial College London}
  \city{London}
  \country{UK}
}

\author{George A. Constantinides}
\email{g.constantinides@imperial.ac.uk}
\affiliation{%
  \institution{Imperial College London}
  \city{London}
  \country{UK}
}

\begin{abstract}

Lookup tables (LUTs) are frequently used to efficiently store arrays of precomputed values for complex mathematical computations. When used in the context of neural networks, these functions exhibit a lack of recognizable patterns which presents an unusual challenge for conventional logic synthesis techniques. Several approaches are known to break down a single large lookup table into multiple smaller ones that can be recombined. Traditional methods, such as plain tabulation, piecewise linear approximation, and multipartite table methods, often yield inefficient hardware solutions when applied to LUT-based NNs.

This paper introduces ReducedLUT, a novel method to reduce the footprint of the LUTs by injecting don't cares into the compression process. This additional freedom introduces more self-similarities which can be exploited using known decomposition techniques. We then demonstrate a particular application to machine learning; by replacing unobserved patterns within the training data of neural network models with don't cares, we enable greater compression with minimal model accuracy degradation. In practice, we achieve up to $1.63\times$ reduction in Physical LUT utilization, with a test accuracy drop of no more than $0.01$ accuracy points. 

\end{abstract}



  
\maketitle

\section{Introduction}
Hardware engineers frequently implement functions using Logical Lookup Tables (L-LUTs), that may exceed the size of an FPGA's Physical Lookup Tables (P-LUTs), in which case they are implemented as a circuit of multiple P-LUTs. For large, user-specified L-LUTs, specialized compression techniques have been developed to provide efficient implementations, beyond those reachable using traditional mapping approaches~\cite{Brayton2010,Soeken2022}. Methods like TwoTable~\cite{twotable} and Lossless Differential Table Compression (LDTC)~\cite{ldtc} have proven effective in reducing hardware utilization by identifying redundancies and exploiting self-similarities within the data. These methods focus on breaking down large L-LUTs into smaller components, thereby minimizing the required resources, particularly in applications such as elementary function approximation. Whilst such compression schemes are effective for L-LUTs with regular structure, recent applications mapping deep neural networks (DNNs) to L-LUTs have introduced a new class of L-LUTs that do not exhibit exploitable properties such as monotonicity or symmetry. Such L-LUTs have remained resistant to effective compression posing a new challenge for efficient implementation~\cite{yukio}. Fortunately, since these irregular L-LUTs implement a DNN, they have an inherent error tolerance, due to over-parameterization of the network. In this paper, we exploit this tolerance to significantly improve L-LUT compression for LUT-based DNNs.

An under-explored yet promising approach to enhancing L-LUT compression involves the strategic incorporation of don't care conditions—input combinations for which the output is either irrelevant or can be approximated without materially affecting the system’s overall functionality or accuracy. Don’t care conditions are particularly prevalent in applications involving approximate computing or machine learning, where exact outputs for all input combinations are not required. This strategy not only has the potential to substantially reduce hardware costs while maintaining or even enhancing system performance, but it also enables the compression of L-LUTs that previously could not be efficiently compressed due to their complex and irregular data patterns. 

In this paper, we present ReducedLUT, a novel method that integrates don’t care conditions into a L-LUT decomposition framework. ReducedLUT increases the self-similarities within L-LUTs during the decomposition process by modifying L-LUT entries associated with these don’t care conditions. The approach leverages the flexibility afforded by the don’t cares, injecting freedom into L-LUT entries where the precise output is non-critical, for example in internal DNN signals. This allows these resistant tables to be restructured in a way that reveals new opportunities for pattern recognition and data redundancy, thus facilitating effective compression—defined as the ratio between the P-LUT utilization using traditional synthesis tools over the P-LUT utilization using our compression technique. This method extends the applicability of L-LUT compression techniques to previously challenging scenarios, enabling the compression of complex tables that were resistant to earlier methods.

To evaluate the effectiveness of ReducedLUT, we apply it to LUT-based implementations of DNNs across two different datasets, targeting a variety of model sizes~\cite{NeuraLUT}. The results demonstrate that with near-zero accuracy drop, ReducedLUT achieves up to {\jscfivelmaxred} reduction in P-LUT utilization.
Furthermore, the method expands the scope of L-LUT compression to a broader range of models and applications.

In summary, the key contributions of this paper are:
\begin{itemize}
    \item We introduce ReducedLUT, an open-source\footnote{https://github.com/ollycassidy13/ReducedLUT} L-LUT compression method that leverages don't care conditions to enhance self-similarities within L-LUTs.
    \item We show that ReducedLUT achieves up to {\jscfivelmaxred} reduction in P-LUT utilization, while preserving the test accuracy with minimal loss, across diverse model sizes and datasets.
    \item We expand the scope of L-LUT compression to include challenging applications like neural networks, where optimizing tables has proven difficult.
\end{itemize}

The paper is organized as follows. Section~\ref{sec:background} provides background on existing compression schemes and LUT-based DNNs. Section ~\ref{sec:motivational} provides a small-scale example that illustrates the objective of our method. In Section~\ref{sec:method}, we describe the ReducedLUT methodology, which we then evaluate on a set of benchmarks in Section~\ref{sec:results}.

\section{Background}\label{sec:background}
\subsection{LUT-based Neural Networks}\label{subsec:lut_dnn}
In the context of neural networks (NNs), efficient P-LUT utilization is critical. Recent advancements in FPGA-based machine learning, such as those demonstrated by LUT-based NNs have shown that L-LUTs are instrumental in implementing computationally intensive functions within NNs. Such NNs have enabled low-latency data classification at CERN and have seen significant advances in recent years~\cite{logicnets,poly,NeuraLUT}. Techniques like LogicNets~\cite{logicnets}, PolyLUT~\cite{poly}, and NeuraLUT\cite{NeuraLUT} have been developed to optimize latency and P-LUT usage for NN inference on FPGAs. LogicNets absorbs the whole computation of a neuron inside an L-LUT. PolyLUT builds on LogicNets by enhancing the representational capacity of NNs by using multivariate polynomial functions inside the neurons, while NeuraLUT further increases NN capacity by absorbing entire sub-networks inside L-LUTs. Despite these innovations, the exponential growth in L-LUT size with increasing input dimensions remains a significant limitation, particularly when larger networks and higher resolutions are required. Furthermore, the L-LUTs generated by LUT-based NNs tend to be irregular, meaning they lack predictable patterns and limit logic simplification. This poses a new challenge for mapping and L-LUT compression schemes.

\subsection{LUT Compression}\label{subsec:lut_compress}

The effectiveness of L-LUTs is inherently constrained by their exponential growth in size with increasing input resolution. For a given function with an input resolution of \( w_{\text{in}} \) and output resolution of \( w_{\text{out}} \), the corresponding L-LUT size scales as \( 2^{w_{\text{in}}} \times w_{\text{out}} \) bits. This exponential scaling quickly renders L-LUTs impractical for high-resolution applications when direct tabulation of all possible inputs is required. To address this challenge, compression techniques have been proposed to reduce the memory footprint of L-LUTs.

Traditional L-LUT compression methods, such as bipartite~\cite{bipartite} and multipartite table decomposition ~\cite{multipartite1, multipartite2}, break a large table down into several smaller tables. These methods decompose a function into a Table of Initial Values and a Table of Offsets to reduce hardware costs. Another approach, piecewise polynomial approximation~\cite{ppa}, divides the function into smaller regions, approximating each with a polynomial whose coefficients are stored in smaller tables. These methods, though effective for functions with smooth or predictable behavior, struggle with highly irregular functions—such as those found in neural networks—where inherent complexities pose significant challenges for compression.

\subsubsection{BDD optimization with don't care conditions}
To address these challenges, recent research has explored incorporating \textit{don’t care} conditions into LUT-based NN tables~\cite{yukio}. A don’t care condition gets attributed to an output that arises from input patterns that are either not observed or are very rare in the training data, permitting freedom in the output for those specific input combinations. In previous work~\cite{yukio}, don’t care conditions were identified and leveraged for minimization of the logic using Binary Decision Diagrams (BDDs), which optimize the logic representation by merging nodes where the output is irrelevant for these inputs. However, neural network functions often produce complex L-LUTs, which are difficult to simplify, even with the help of BDD minimization with the inclusion of don't care conditions. By contrast, we adopt a more direct approach by using truth table decomposition, where we operate directly on the L-LUT implementation to uncover and exploit additional patterns. Unlike BDD minimization, which works at a more abstract level, our method focuses on tailoring the L-LUTs for efficient table decomposition.

\subsubsection{CompressedLUT}
\label{sec:compressedlut}

\begin{figure}[b]
\centerline{\includegraphics[width=70mm]{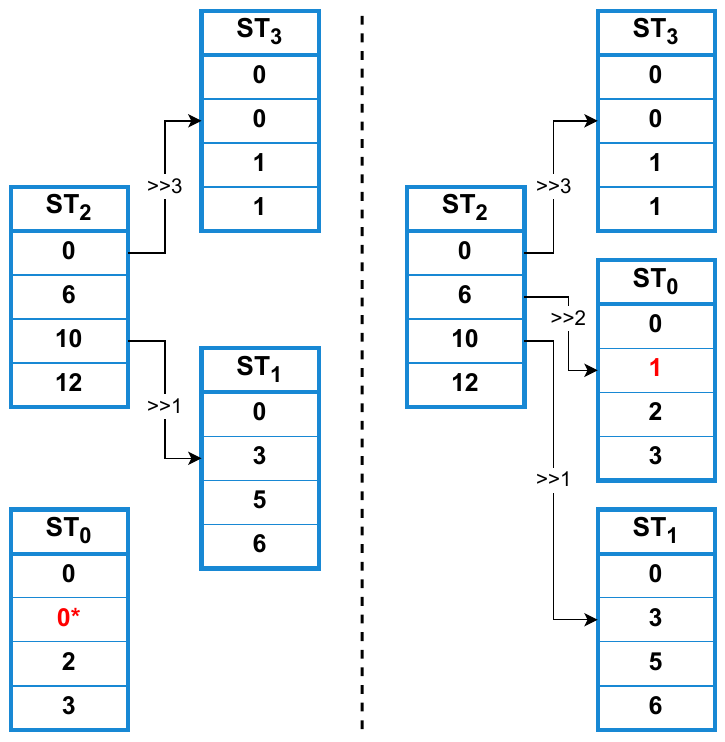}}
\caption{Don't care usage example.}
\label{fig:dont_care_example}
\end{figure}

In this paper, we build upon CompressedLUT, which employs a systematic approach to reducing the storage requirements of L-LUTs whilst preserving data fidelity~\cite{compressedlut}. The process begins with the decomposition of an original table, $T$, into two smaller tables, $T_{\text{bias}}$ and $T_{\text{st}}$. Specifically, $T$ is divided into sub-tables such that the minimum value of each sub-table is stored in $T_{\text{bias}}$ and the residual values, after subtracting the minimum, are stored in $T_{\text{st}}$. This decomposition is particularly effective in reducing the bitwidth required for storing $T_{\text{st}}$ as it captures only the local variations within the data. 

CompressedLUT enhances this methodology by exploiting self-similarities within $T_{\text{st}}$, where self-similarity refers to the presence of repetitive or highly similar patterns across different sub-tables, allowing for data redundancy to be minimized. By constructing a similarity matrix, sub-tables that are either identical or can be derived from one another through arithmetic right shifts are identified. Instead of storing multiple similar sub-tables, CompressedLUT constructs $T_{\text{ust}}$, which only stores unique sub-tables, while the remaining are reconstructed using the index and shift information contained in $T_{\text{idx}}$ and $T_{\text{rsh}}$, respectively. This self-similarity detection significantly reduces the overall memory footprint. After decomposing the input, $x=\{x_{hb},x_{lb}\}$, the original table entry $T[x]$ can be reconstructed using the following:
\begin{equation}
\label{eq:1}
    T[x] = \left( T_{\text{ust}}\left[\left\{ T_{\text{idx}}[x_{hb}], x_{lb} \right\}\right] \gg T_{\text{rsh}}[x_{hb}] \right) + T_{\text{bias}}[x_{hb}].
\end{equation}

This process is repeated, iterating through multiple sub-table lengths in an attempt to find a better compression. To further enhance the efficiency of the compression, particularly for tables with higher dynamic range and large local variations, CompressedLUT introduces a higher-bit compression technique. In this higher-bit compression technique, the table \( T \) is split into a higher-bits table, $T_{hb}$ and a lower bits table, $T_{lb}$, which are then concatenated to form the output, $T[x] = \{T_{hb}[x], T_{lb}[x]\}$. The lower bits table $T_{lb}$ is not compressed, while the higher bits undergo decomposition and self-similarity detection as shown in~\eqref{eq:1}.

CompressedLUT starts with no higher-bit compression then iterates through different splitting configurations to find the optimal solution. These techniques are effective when table values change continuously but face issues when presented with small local variations. By splitting $T$ into $T_{hb}$ and $T_{lb}$ before compression, the technique reduces local variations when searching for self-similarities, improving compression efficiency.

Whilst usually effective, neural network functions tend to appear random-looking, further complicating L-LUT decomposition techniques such as CompressedLUT, which rely on pattern recognition for optimization. Therefore, we develop a CompressedLUT-based method that leverages don't cares to exhibit more self-similarities within a table. Specifically, we aim to reduce the number of unique sub-tables, $T_{\text{ust}}$, by matching more values to previously stored sub-tables. ReducedLUT is a domain-specific enhancement of CompressedLUT, designed to further optimize the decomposition process by leveraging unobserved patterns within the NN training data. Further details about the methodology can be found in Section~\ref{sec:method}.

\section{Motivational Example} \label{sec:motivational}
To motivate the incorporation of don't cares into L-LUT compression, we show an example in Figure~\ref{fig:dont_care_example}. Following the CompressedLUT methodology~\cite{compressedlut}, the initial table, $T$, has been decomposed into four smaller sub-tables, $ST_0$, $ST_1$, $ST_2$ and $ST_3$ and a bias table $T_{\text{bias}}$ (not shown). Without modification, we can only reconstruct $ST_1$ and $ST_3$ from $ST_2$ by applying right-shifts of one and three, respectively. We now say that $ST_1$ and $ST_3$ depend on $ST_2$. However, in $ST_0$, one entry is marked as a don't care. By replacing the second entry of $ST_0$, with a one, it is now possible to reconstruct $ST_0$ from $ST_2$ by applying a right-shift by two. $ST_0$ now also depends on $ST_2$. Having shown that modifying don't care values can improve compression, we now describe ReducedLUT in detail.

\begin{figure*}
    \centering
    \includegraphics[width=\linewidth]{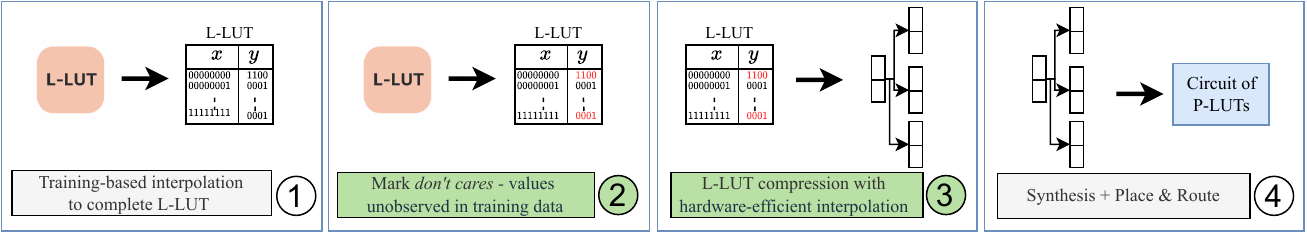}
    \caption{ReducedLUT problem formulation and toolflow.}
    \label{fig:toolflow}
\end{figure*}

\section{ReducedLUT Methodology} \label{sec:method}

The ReducedLUT toolflow is shown in Figure~\ref{fig:toolflow}. In this paper, we take as input a trained network of L-LUTs. As in prior work~\cite{NeuraLUT}, the content of each L-LUT is derived from an interpolation of the training data performed by the functional form used in training. For each L-LUT within this network, ReducedLUT identifies the values unobserved in the training data and marks those L-LUT entries as don't cares. The key idea behind ReducedLUT is to replace some of these interpolation values determined during NN training with values obtained via hardware-efficient interpolation during the L-LUT compression process. As in Section~\ref{sec:motivational}, the replacement is performed in order to maximize the number of self-similarities between sub-tables, facilitating a reduction in the number of unique sub-tables that need to be implemented. We now provide a detailed description of each step. 

\subsection{Identification of Don't Care Conditions}\label{subsec:identify}
In LUT-based NNs, we evaluate all the possible L-LUT input combinations to generate completely specified L-LUTs. To identify don't care entries, ReducedLUT performs an inference using the training dataset, and marks unobserved L-LUT inputs as don't cares. For example, in Figure~\ref{fig:toolflow}, we never observe the input \texttt{00000000} as an input to the given L-LUT, therefore we class its output as a don't care. Prior work on LUT-based NNs, has shown that assigning arbitrary values to these unobserved L-LUT entries does not significantly degrade the test accuracy of the overall NN~\cite{yukio}. The entries classified as don't cares may later be replaced with alternative concrete values in the ReducedLUT flow as we aim for a lower cost of implementing that neuron's function. Moreover, since our method only marks the unobserved entries in the training dataset, the training accuracy will be unchanged by the L-LUT compression and we aim to ensure the function still generalizes well by preserving the general circuit structure.

The ReducedLUT strategy can be formally described as follows. Let $f:X\to Y$ denote the function computed by the pre-trained NN, mapping from the input domain $X$ to the output domain $Y$. Given a training data set, $X_{\textrm{train}}\subseteq X$, ReducedLUT constructs $\tilde{f}: X \to Y$ to minimize a hardware area cost function. $\tilde{f}$ is selected from a set of candidate functions $\mathcal{F}$ that leverage the structural properties of $f$.

\begin{align}
\min_{\tilde{f} \in \mathcal{F} } \quad & \text{area\_cost}(\tilde{f}) \\
\text{subject to} \quad & \tilde{f}(x_i) = f(x_i), \quad \forall x_i\in X_{\textrm{train}}
\end{align}

\subsection{L-LUT Compression}\label{subsec:compress}
As in CompressedLUT, each L-LUT, $T$, is first decomposed into a set of sub-tables $T_{st} = \{ST_1, \dots, ST_n\}$ each containing $M$ entries. The algorithm examines the sub-table’s ability to generate other sub-tables, $ST_j$, through bitwise right shifts in order to compute a binary similarity matrix $SM$. Each entry $SM_{i,j}$ indicates whether $ST_j$ can be reconstructed from $ST_i$ via a right-shift, which we store in a \textit{right-shift} matrix, $SM_{i,j}^{rsh}$. More precisely,
\begin{equation}
\begin{split}
    SM_{i,j} = 1 \text{  and  }  SM_{i,j}^{rsh} = t \iff  \\
    \forall k, ST_i[k] \gg t = ST_j[k]
\end{split}
\end{equation}

The sub-tables are ranked using the sum of each column of the similarity matrix to form the similarity vector, $SV$, where the values in the similarity vector represent the number of dependencies each sub-table has~\cite{compressedlut}. 

ReducedLUT first treats every entry as a \textit{care}, identifying all the unique sub-tables and storing their indices in a vector $I_{ust}$. This is performed by initially selecting the first unique sub-table, $ST_i$, based on the index of the maximum value in $SV$. The index corresponding to this sub-table, $i$, is stored as the first unique index in $I_{ust}$. If the sub-table can generate others, the similarity matrix is then updated by setting the $i\text{th}$ row and column to zero. The corresponding row and column of the similarity matrix are also zeroed for each sub-table $ST_j$ that can be generated from $ST_i$. The similarity vector is finally re-calculated based on this updated matrix. This iterative process continues until all entries in the similarity matrix are zero, indicating that all unique sub-tables have been identified.

ReducedLUT aims to reduce the number of unique sub-tables by attempting to match sub-tables to each other by changing values that are marked as don't cares. Each time a unique sub-table entry is modified, the set of sub-tables that can be derived from it is also modified, potentially breaking existing dependencies. Each time a dependency is broken the corresponding sub-table(s) must be re-allocated to an alternative unique sub-table or added to the unique sub-table list. Since it is typically simpler to modify unique sub-tables with fewer dependencies, $I_{ust}$ is sorted based on the capacity of each sub-table to generate others which we get from the similarity vector. 

The sorting is performed because ReducedLUT starts with the sub-table with the fewest dependencies, call it $ST_{\textrm{min}}$, and attempts to modify its don't care entries, such that it can be generated from another unique sub-table, beginning the search with the unique sub-table with the greatest number of dependencies. If ReducedLUT successfully replaces the don't cares such that $ST_{\textrm{min}}$ can be generated from another unique sub-table, it must also check whether all of its dependencies can be re-generated from the remaining unique sub-tables. This re-generation process can further exploit the don't cares in sub-tables that depend on $ST_{\textrm{min}}$. Any potential changes are implemented provisionally, with the original values and corresponding indices temporarily stored to enable rollback if the modification does not result in a reduction in $I_{ust}$. This backtracking search helps identify more efficient decomposition structures without significantly increasing the runtime. ReducedLUT iterates over the entire $I_{ust}$ list attempting to modify each unique sub-table in-turn. 

A Boolean mask is employed to track the modifiability of each entry within the sub-tables, ensuring that certain entries remain fixed once they have been used in successful transformations. This mechanism is crucial for maintaining the integrity of the compression process by preventing further alterations to sub-tables that have already contributed to a successful match, thereby preserving previous optimizations and ensuring consistent progression throughout the compression.

Finally, the entire optimization algorithm is repeated for various sub-table sizes (values of $M$) and for different higher-bit compression configurations as per Section~\ref{sec:compressedlut}. After the L-LUT compression algorithm with don't care integration is complete, ReducedLUT constructs the final decomposed table which is then output in Verilog.

\begin{table*}[t]
\renewcommand{\arraystretch}{1.2} 
\centering
\caption{NeuraLUT model architectures used for evaluation. $\beta_0$ and $F_0$ are the input layer's activation bitwidth and fan-in.}
\label{table:neuralut}
\begin{tabular}{|l|l|c|c|c|c|c|c|c|c|}
\hline
\textbf{Dataset} & \textbf{Model Name} & \textbf{L-LUTs per Layer} & $\beta$ & \textbf{F} & $\beta_0$ & $\textbf{F}_0$ & \textbf{Test Accuracy} \\ \hline
MNIST           & MNIST              & 256, 100, 100, 100, 10    & 2       & 6   & 2       & 6       & 95.760\%              \\ \hline
\multirow{2}{*}{Jet substructure} 
                            & JSC-2L & 32, 5 & 4 & 3 & 4 & 3 & 72.525\%     \\ \cline{2-8} 
                            & JSC-5L & 128, 128, 128, 64, 5 & 4 & 3 & 7 & 2 & 74.906\%  \\ \hline
\end{tabular}
\end{table*}

\subsection{Exiguity}

During the process of L-LUT compression, we introduce a parameter called \textit{exiguity}, which sets a threshold on how many sub-tables can depend on a given unique sub-table for it to be eligible for don't care optimization. If a unique sub-table is altered, all the sub-tables that depend on it must also be modified to be generated from other unique sub-tables; otherwise, there would be no reduction in the total number of unique sub-tables. When a unique sub-table has too many dependencies, it becomes increasingly difficult to alter all of them. If this cannot be achieved, the changes are rolled back, inefficiently increasing runtime without achieving any further compression.

In practice, \textit{exiguity} ensures that the algorithm prioritizes sub-tables with fewer dependencies, preventing excessive computational overhead and long runtimes. By limiting the number of dependencies a sub-table can have to be eligible for modification, exiguity helps achieve not only better compression results but also almost no loss in test accuracy. This comes as a result of preventing the program from touching patterns that are critical for preserving the original L-LUT structure.

\section{Experimental Results} \label{sec:results}
To evaluate the accuracy of the ReducedLUT generated NNs, we perform an inference on the Verilog circuit using the test data to get our final test accuracy. The final 6-input P-LUT utilization is obtained through Vivado $2020.1$ synthesis, selecting the \texttt{xcvu9p-flgb2104-2-i} FPGA part, to allow direct comparison to NeuraLUT~\cite{NeuraLUT}.

\subsection{Benchmarks}
\begin{table}
\centering
\renewcommand{\arraystretch}{1.2}
\caption{ReducedLUT evaluation at different exiguity levels.}
\setlength{\tabcolsep}{2pt} 
\renewcommand{\arraystretch}{1.3} 
\begin{tabular}{|c|c|c|c|c|c|}
\hline
\textbf{Dataset} & \textbf{Method} & \textbf{Exiguity} & \textbf{P-LUTs} & \shortstack{\rule{0pt}{8pt} \textbf{$F_{\text{max}}$} \\ \textbf{(MHz)}} & \shortstack{\textbf{Test} \\ \textbf{Accuracy}} \\ \hline \hline

\multirow{6}{*}{JSC\_2L} 
    & Baseline      & -    & 4387  & 755.3 & 72.525\% \\ \cline{2-6}
    & CompressedLUT   & -   & 3299  & 333.4 & 72.525\% \\ \cline{2-6} 
    & Random Values & - & - & - & 72.494\% \\ \cline{2-6}
    & \multirow{3}{*}{\textbf{ReducedLUT}} 
        & 20   & 3071  & 364.8 & 72.525\% \\ \cline{3-6}
        && 150  & 2895  & 390.0 & 72.525\% \\ \cline{3-6} 
        && 250  & \textbf{2786} & 408.5 & 72.525\% \\ \hline \hline 
\multirow{6}{*}{JSC\_5L} 
    & Baseline      & -    & 95251 & 329.8 & 74.906\% \\ \cline{2-6} 
    & CompressedLUT   & -   & 61278 & 272.1 & 74.906\% \\ \cline{2-6}
    & Random Values & - & - & - & 74.746\% \\ \cline{2-6}
    & \multirow{3}{*}{\textbf{ReducedLUT}} 
        & 20   & 59211 & 269.5 & 74.907\% \\ \cline{3-6}
        && 150  & 58899 & 294.8 & 74.904\% \\ \cline{3-6}
        && 250  & \textbf{58409} & 302.8 & 74.904\% \\ \hline \hline
\multirow{6}{*}{MNIST} 
    & Baseline      & -    & 53548 & 302.2 & 95.760\% \\ \cline{2-6}
    & CompressedLUT   & -   & 51831 & 291.4 & 95.760\% \\ \cline{2-6}
    & Random Values & - & - & - & 95.417\% \\ \cline{2-6}
    & \multirow{3}{*}{\textbf{ReducedLUT}} 
        & 20   & 49001 & 291.9 & 95.760\% \\ \cline{3-6}
        && 150  & 49336 & 292.7 & 95.760\% \\ \cline{3-6}
        && 250  & \textbf{47484} & 295.4 & 95.750\% \\ \hline

\end{tabular}

\label{tab:benchmarks}
\end{table}
We utilized models from the NeuraLUT project \cite{NeuraLUT} as the basis for our study. These models are designed as quantized sparse neural networks, where each neuron is implemented as an L-LUT, forming a network of truth tables. Here, $\beta$ indicates the bit-width of each activation, and $F$ represents the fan-in, or the number of input activations per neuron, which remains consistent throughout the hidden layers of the network. For each truth table, the input bit-width is calculated as $\beta \times F$, while the output bit-width is $\beta$. Table~\ref{table:neuralut} provides the specifications of the neural networks utilized in our experiments. The NeuraLUT networks were trained using two publicly accessible datasets: 1) jet substructure classification (JSC) with a train/test split of 800k/200k; and 2) handwritten digit classification (MNIST) with a train/test split of 60k/10k.

To evaluate the advantage of injecting don't cares, we apply both the original CompressedLUT~\cite{compressedlut}, using the open-source code\footnote{\url{https://github.com/kiabuzz/CompressedLUT/tree/main}}, and ReducedLUT to these NeuraLUT networks of L-LUTs. To demonstrate how ReducedLUT preserves accuracy, we also include a naive approach, where the don’t care values are replaced with randomly generated values\footnote{\url{https://github.com/ollycassidy13/RandomInsertion}}.

\subsection{Optimisation Results}
In Table~\ref{tab:benchmarks}, we present results on three benchmarks—JSC\_2L, JSC\_5L, and MNIST. The Baseline results are collected using the NeuraLUT toolflow, which directly passes the L-LUTs to the Vivado synthesis tool. We report results for CompressedLUT and ReducedLUT with varying exiguity values. For all benchmarks and exiguity values tested, ReducedLUT significantly decreased P-LUT utilization compared to the baseline, without degrading the overall neural network's test accuracy. Overall, the results consistently demonstrate that ReducedLUT surpasses CompressedLUT, giving an 8.7\% geometric mean reduction in P-LUT utilization, underscoring the effectiveness of our approach in leveraging don’t care conditions within the L-LUT compression framework. 

For the JSC\_2L and JSC\_5L models, ReducedLUT decreased P-LUT utilization by up to {\jsctwolmaxred} and {\jscfivelmaxred} when compared to the baselines, with almost no degradation in test accuracy. For the MNIST model, P-LUT utilization was only reduced by up to {\mnistmaxred}. This smaller reduction stems from the fact that MNIST’s L-LUT output bitwidth is limited to two, which restricts the ability of both CompressedLUT and ReducedLUT to separate local variations into distinct sub-tables. As a result, the sub-tables exhibit more local variations, reducing self-similarities and generally making compression less effective. This is a limitation of the CompressedLUT methodology, which suffers in scenarios with few distinct sub-table patterns and high local variation. While ReducedLUT improves compression by leveraging don't care conditions, the inherent lack of self-similarities in these sub-tables limits further compression gains for MNIST. ReducedLUT mitigates some of the increased latency introduced by CompressedLUT, as its simplified logic alleviates timing pressure during the synthesis.

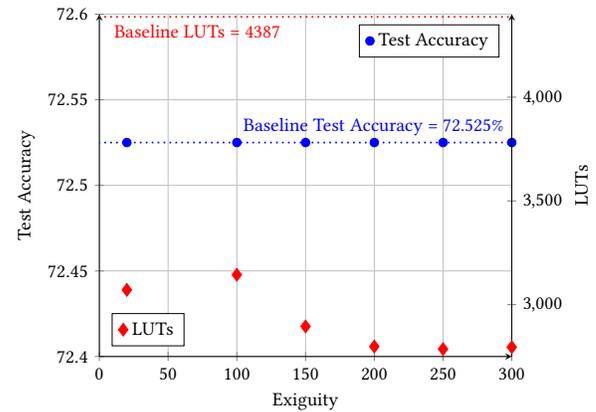
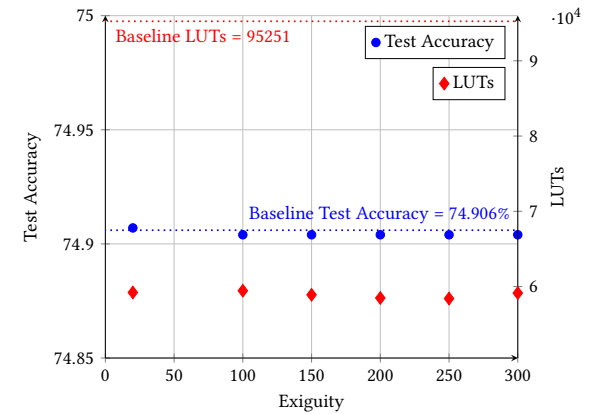
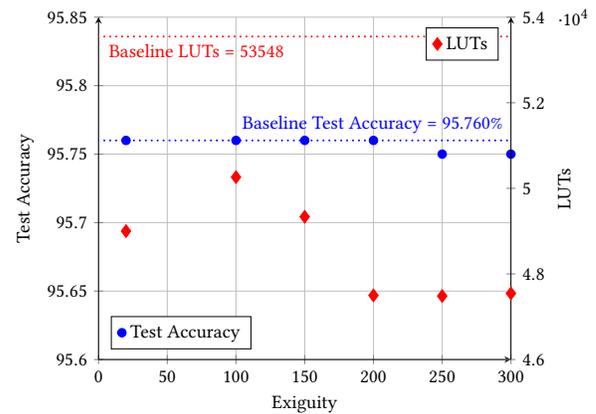
\begin{figure}[ht]
    \centering
    \begin{subfigure}{0.48\textwidth}
        \centering
        \begin{tikzpicture}[scale=0.8]  
        \begin{axis}[
            xlabel={Exiguity},
            ylabel={Test Accuracy},
            legend pos=north east,
            grid=major,
            axis y line=left,
            axis x line=bottom,
            xmin=0, xmax=300,
            ymin=72.4, ymax=72.6
        ]
    
        \draw[dotted, thick, blue] (axis cs:-5,72.525) -- (axis cs:300,72.525)
        node[pos=0.67, anchor=south, color=blue] {Baseline Test Accuracy = 72.525\%};  
    
        \addplot[
            only marks,
            mark=*,
            color=blue
        ]
        coordinates {
        (20,72.525)
        (100,72.525)
        (150,72.525)
        (200,72.525)
        (250,72.525)
        (300,72.525)
        };
        \addlegendentry{Test Accuracy}
    
        \end{axis}
    
        \begin{axis}[
            xlabel={Exiguity},
            ylabel={LUTs},
            axis y line=right,
            axis x line=none,
            xmin=0, xmax=300,
            ymin=2750, ymax=4400,
            legend pos=south west
        ]
    
        \draw[dotted, thick, red] (axis cs:-5,4387) -- (axis cs:300,4387)
        node[pos=0.25, anchor=north, color=red] {Baseline LUTs = 4387};  
    
        \addplot[
            only marks,
            mark=diamond*,
            mark size=3,
            color=red
        ]
        coordinates {
        (20,3071)
        (100,3144)
        (150,2895)
        (200,2798)
        (250,2786)
        (300,2795)
        };
        \addlegendentry{LUTs}
    
        \end{axis}
        \end{tikzpicture}
        \caption{Test Accuracy and P-LUTs vs Exiguity for the JSC-2L model.}
    \end{subfigure}
    \hfill
    \begin{subfigure}{0.48\textwidth}
        \centering
        \begin{tikzpicture}[scale=0.8]  
        \begin{axis}[
            xlabel={Exiguity},
            ylabel={Test Accuracy},
            legend pos=north east,
            grid=major,
            axis y line=left,
            axis x line=bottom,
            xmin=0, xmax=300,
            ymin=74.85, ymax=75
        ]

        \draw[dotted, thick, blue] (axis cs:-5,74.906) -- (axis cs:300,74.906)
        node[pos=0.67, anchor=south, color=blue] {Baseline Test Accuracy = 74.906\%};  
        
        \addplot[
            only marks,
            mark=*,
            color=blue
        ]
        coordinates {
        (20,74.907)
        (100,74.904)
        (150,74.904)
        (200,74.904)
        (250,74.904)
        (300,74.904)
        };
        \addlegendentry{Test Accuracy}

        \end{axis}

        \begin{axis}[
            xlabel={Exiguity},
            ylabel={LUTs},
            axis y line=right,
            axis x line=none,
            xmin=0, xmax=300,
            ymin=50500, ymax=96000,
            legend style={at={(0.97,0.85)}}
        ]

        \draw[dotted, thick, red] (axis cs:-5,95251) -- (axis cs:300,95251)
        node[pos=0.25, anchor=north, color=red] {Baseline LUTs = 95251};  
        
        \addplot[
            only marks,
            mark=diamond*,
            mark size=3,
            color=red
        ]
        coordinates {
        (20,59211)
        (100,59446)
        (150,58899)
        (200,58481)
        (250,58409)
        (300,59133)
        };
        \addlegendentry{LUTs}

        \end{axis}
        \end{tikzpicture}
        \caption{Test Accuracy and P-LUTs vs Exiguity for the JSC-5L model.}
    \end{subfigure}
    \hfill
    \begin{subfigure}{0.48\textwidth}
        \centering
        \begin{tikzpicture}[scale=0.8]  
        \begin{axis}[
            xlabel={Exiguity},
            ylabel={Test Accuracy},
            legend pos=south west,
            grid=major,
            axis y line=left,
            axis x line=bottom,
            xmin=0, xmax=300,
            ymin=95.6, ymax=95.85
        ]

        \draw[dotted, thick, blue] (axis cs:-5,95.760) -- (axis cs:300,95.760)
        node[pos=0.67, anchor=south, color=blue] {Baseline Test Accuracy = 95.760\%};  
        
        \addplot[
            only marks,
            mark=*,
            color=blue
        ]
        coordinates {
        (20,95.76)
        (100,95.76)
        (150,95.76)
        (200,95.76)
        (250,95.75)
        (300,95.75)
        };
        \addlegendentry{Test Accuracy}

        \end{axis}

        \begin{axis}[
            xlabel={Exiguity},
            ylabel={LUTs},
            axis y line=right,
            axis x line=none,
            xmin=0, xmax=300,
            ymin=46000, ymax=54000,
            legend pos=north east
        ]

        \draw[dotted, thick, red] (axis cs:-5,53548) -- (axis cs:300,53548)
        node[pos=0.25, anchor=north, color=red] {Baseline LUTs = 53548};  
        
        \addplot[
            only marks,
            mark=diamond*,
            mark size=3,
            color=red
        ]
        coordinates {
        (20,49001)
        (100,50263)
        (150,49336)
        (200,47499)
        (250,47484)
        (300,47546)
        };
        \addlegendentry{LUTs}

        \end{axis}
        \end{tikzpicture}
        \caption{Test Accuracy and P-LUTs vs Exiguity for the MNIST model.}
    \end{subfigure}
    \caption{Test accuracy and P-LUT utilization for different exiguity levels against the NeuraLUT baseline.} 
    \label{fig:exiguity graphs}
\end{figure}

The results in Figure~\ref{fig:exiguity graphs} show that larger exiguity values generally result in greater reductions in P-LUT utilization, by decreasing the number of unique sub-tables required to reconstruct the L-LUTs. We also observe a saturation in compression as the exiguity increases. For the JSC\_5L model, the initial drop from the baseline is quite significant, indicating that the set of unique sub-tables is already relatively compact, and each sub-table contains a large number of dependencies. This makes it difficult to further improve the compression ratio, as reallocating the dependencies associated with any unique sub-table becomes increasingly challenging.

When replacing don't cares with concrete values, our algorithm does so on the basis of what is likely to improve compression, and pays no attention to the potential impact on the generalization performance of the network. It is therefore possible that compression for area reduction and generalization using our method are in tension. The randomization results in Table~\ref{tab:benchmarks} show this not to be the case: decisions taken by our algorithm to compress the circuits correspond, in general, to decisions that do not impact the accuracy of the result, even though random decisions {\em would} impact this accuracy. One interpretation of this result is that our technique naturally makes use of the generalization inherent in compression, an idea known as Occam's Razor~\cite{computational_learning1994}.

Both ReducedLUT and CompressedLUT were run on an Intel Xeon E5 processor. Across all the benchmarks, ReducedLUT's average runtime per L-LUT input is 1.37 times larger than CompressedLUT at 250 exiguity.  
Increasing the exiguity parameter from 20 to 250, ReducedLUT took 19\% longer to execute.

\section{Conclusion and Future Work}
\label{sec:furtherwork}
In this paper, we introduced ReducedLUT, a method to compress LUT-based NNs through the strategic replacement of don't care table entries. For a given LUT-based NN, ReducedLUT identifies L-LUT values that can be replaced without affecting the network's training accuracy. By then exploiting don't care entries in the L-LUT compression, ReducedLUT shrinks P-LUT usage of LUT-based NNs by up to {\jscfivelmaxred}, with minimal impact on the testing accuracy. 

Future work will extend the don't care set beyond values that do not occur in the training dataset to also include values that occur infrequently. Incorporating these rare values into the don't care set could further optimize L-LUT compression without significantly sacrificing accuracy, as these values have minimal impact on the overall network performance. Additionally, compression could be computed for groups of L-LUTs, rather than treating each L-LUT independently. This would enhance L-LUT sharing across the network, potentially leading to even greater reductions in L-LUT utilization and improving hardware efficiency.

\bibliographystyle{ACM-Reference-Format}
\bibliography{bibliography}

\end{document}